\documentclass[10pt,twocolumn,letterpaper]{article}

\usepackage[pagenumbers]{cvpr} 
\usepackage[accsupp]{axessibility}  

\usepackage{graphicx}
\usepackage{amsmath}
\usepackage{amssymb}
\usepackage{pifont}
\usepackage{booktabs}
\usepackage{widetable}
\usepackage{bm}
\usepackage{color}
\usepackage{tabularx}
\usepackage{mathtools}
\usepackage[accsupp]{axessibility} 

\newcommand{\cmark}{\ding{51}}
\newcommand{\xmark}{\ding{55}}

\usepackage[pagebackref,breaklinks,colorlinks]{hyperref}

\usepackage[capitalize]{cleveref}
\crefname{section}{Sec.}{Secs.}
\Crefname{section}{Section}{Sections}
\Crefname{table}{Table}{Tables}
\crefname{table}{Tab.}{Tabs.}

\def\cvprPaperID{3} 
\def\confName{CVFAD}
\def\confYear{2024}

\newcommand{\argmin}{{\rm arg} \mathop{\rm min}\limits}

\begin{document}

\title{A Fashion Item Recommendation Model in Hyperbolic Space}
\author{
\textnormal{Ryotaro Shimizu}$^{1,2}$ \textnormal{,} \hspace{5mm} \textnormal{Yu Wang}$^{1}$ \textnormal{,} \hspace{5mm} \textnormal{Masanari Kimura}$^{3}$ \textnormal{,} \\
\textnormal{Yuki Hirakawa}$^{2}$ \textnormal{,} \hspace{5mm} \textnormal{Takashi Wada}$^{2}$ \textnormal{,} \hspace{5mm} \textnormal{Yuki Saito}$^{2}$ \textnormal{,} \hspace{5mm} \textnormal{Julian McAuley}$^{1}$ \\
{$^1$University of California, San Diego \hspace{5mm} $^2$~ZOZO Research
\hspace{5mm} $^3$~The University of Melbourne}
\\
{\tt\small \{r2shimizu, yuw164\}@ucsd.edu}, {\tt\small m.kimura@unimelb.edu.au}, \\
{\tt\small \{yuki.hirakawa,takashi.wada,yuki.saito\}@zozo.com}, 
{\tt\small jmcauley@ucsd.edu}
}

\maketitle

\begin{abstract}
\vspace{-0mm}
In this work, we propose a fashion item recommendation model that incorporates hyperbolic geometry into user and item representations. Using hyperbolic space, our model aims to capture implicit hierarchies among items based on their visual data and users' purchase history. During training, we apply a multi-task learning framework that considers both hyperbolic and Euclidean distances in the loss function. Our experiments on three data sets show that our model performs better than previous models trained in Euclidean space only, confirming the effectiveness of our model. Our ablation studies show that multi-task learning plays a key role, and removing the Euclidean loss substantially deteriorates the model performance.\footnote{This work was presented at the CVFAD Workshop at CVPR 2024.}
\end{abstract}

\vspace{-5mm}
\section{Introduction}
\label{sec:intro}
\vspace{-0mm}
In hyperbolic space, the distance from the origin increases exponentially as one moves towards the outside of the space.
This property makes it well-suited for modeling hierarchical data by representing the root node around the origin and leaf nodes near the surface of the space.
In practice, hyperbolic space has been used to represent various types of instances that potentially possess hierarchical structures, such as sentences and words~\cite{hyperbolic_word1,hyperbolic_word2}, questions and answers~\cite{hyperbolic_qa}, and objects and scenes~\cite{Ge_2023_CVPR}.
In computer vision, its effectiveness has been confirmed on many tasks~\cite{hyperbolic_cv_survey,Hyperbolic_Survey}, including classification~\cite{hyperbolic_zero_shot_recognition_2020_cvpr,hyperbolic_image_text_representations,hyperbolic_image_classificaion_cvr_2021}, retrieval~\cite{hyperbolic_image_retrieval_2024_wacv,hyperbolic_image_retrieval_2021_cvpr,hyperbolic_internet_image_retrieval, hyperbolic_medical_image_retrieval}, segmentation~\cite{Ge_2023_CVPR,hyperbolic_image_segmentation_1,hyperbolic_image_segmentation_2}, and generation~\cite{hyperbolic_image_generation_iccv_2023}.

\vspace{-0mm}
\begin{figure}[t]
\centering
\includegraphics[width=1.\linewidth]{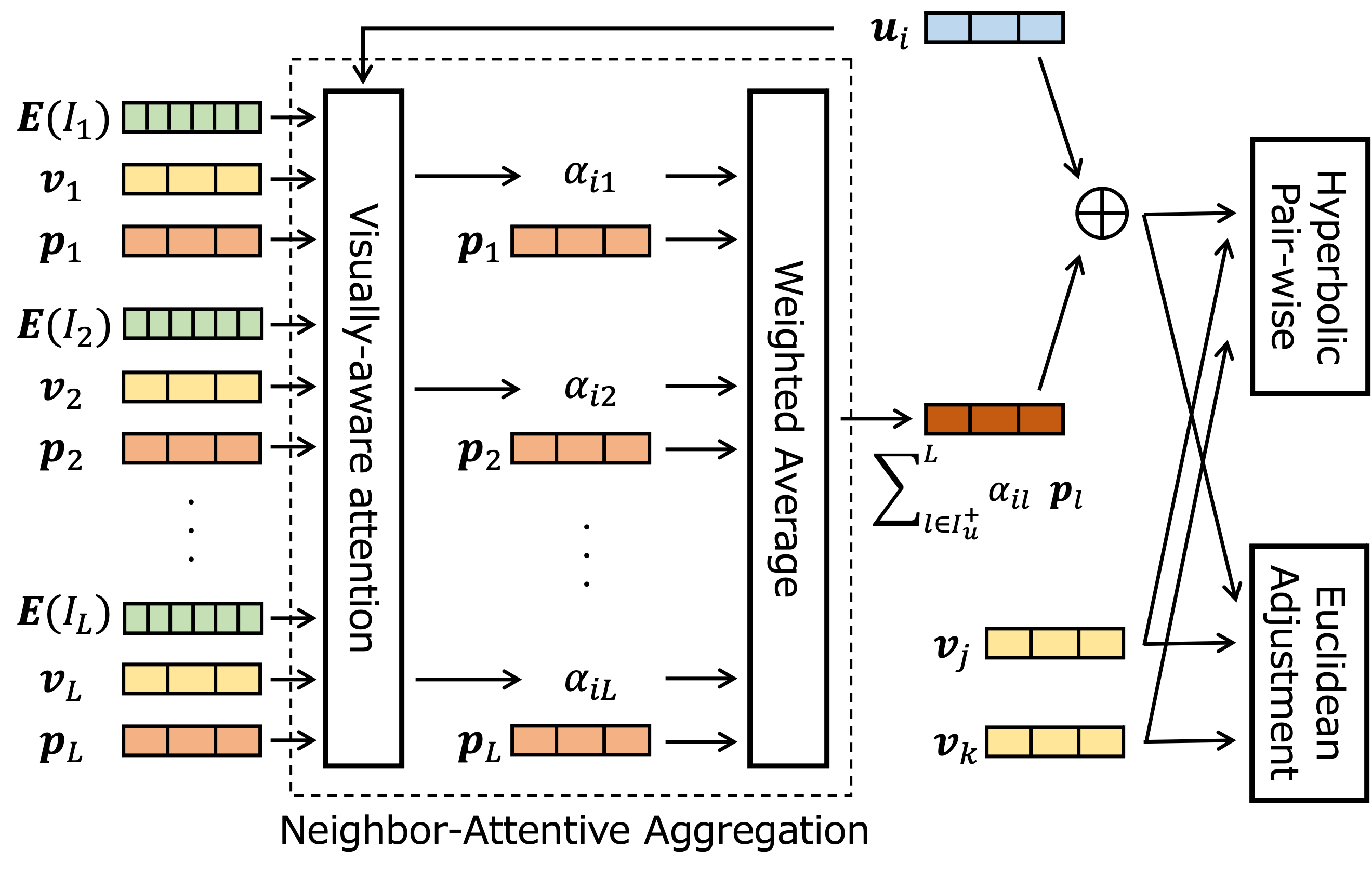}
\vspace{-5mm}
\caption{Overview of our proposed model. \label{fig_structure}}
\end{figure}

In this paper, we propose a new fashion item recommendation model that incorporates hyperbolic geometry into user and item representations; we name it Hyperbolic Visual Attentive Collaborative Filtering (HVACF).
Using hyperbolic space, our model aims to capture implicit hierarchies among fashion items based on their image data and users' purchase history.
To train our model, we apply a multi-task learning framework that combines the hyperbolic distance with the Euclidean one, which was found effective in previous studies for training recommender systems without visual features and also for learning representations of objects and scenes~\cite{Tran_HyperML, Ge_2023_CVPR}.
Our experiments show that our model outperforms existing models trained in Euclidean space only, confirming the effectiveness of our model.
We also conduct ablation studies and find that multi-task learning plays a key role, and removing the Euclidean loss results in poor performance, even worse than baseline systems without visual features.
This result suggests that multi-task learning is crucial for training fashion item recommender systems, in contrast to the previous findings on recommendation tasks without visual features that hyperbolic models perform better than the Euclidean counterparts~\cite{Tran_HyperBPR,hyper_rec_others_1}.

\vspace{-0mm}
\section{Related Work}
\label{sec:preliminaries}
\vspace{-0mm}
\subsection{Visually-Aware Recommender Systems}
Visually-aware recommendation models leverage item visual data to capture characteristics of items and users~\cite{McAuley_VBPR,Kang2017_DVBPR,DeepStyle,ACF}.
Visual information is particularly valuable for recommending items whose appearances and styles heavily affect users' preferences, such as clothing~\cite{Kang2017_DVBPR,DeepStyle,Shimizu2023_FIS}, cosmetic products~\cite{cosme_recsys1,cosme_recsys2}, and hair-style~\cite{hairstyle_recsys1,hairstyle_recsys2}~\cite{McAuley2023_fashionrec_review}.
In this work, we incorporate hyperbolic geometry into a fashion item recommender system with visual features and demonstrate its effectiveness.

\vspace{-0mm}
\subsection{Recommender Systems in Hyperbolic Space}
Hyperbolic geometry differs from Euclidean geometry in that it measures distances on curves rather than straight lines.
In hyperbolic space, the distance from the origin increases exponentially as one moves towards the surface of the space.
This property is well-suited for modeling hierarchies and has been used for representation learning on various tasks~\cite{hyperbolic_embeddings_1,hyperbolic_embeddings_2,hyperbolic_embeddings_3,hyperbolic_embeddings_4,hyperbolic_embeddings_5,hyperbolic_embeddings_6,hyperbolic_embeddings_7,hyperbolic_embeddings_8,hyperbolic_embeddings_9}.
Previous work also showed that it is effective in capturing the complex interactions between items and user preferences in recommender systems~\cite{Hyperbolic_Survey,Tran_HyperML,HyperMF_music,hyper_rec_others_1,hyper_rec_others_2}.
For instance, Tran et al.~\cite{Tran_HyperBPR} proposed a method based on hyperbolic space, called Hyperbolic Bayesian Personalized Ranking (BPR), and Benjamin et al.~\cite{hyper_rec_others_1} extended this model using large-scale data.
Lucas et al.~\cite{Tran_HyperML} proposed Hyperbolic Metric Learning (HML), which incorporates the hyperbolic distance into metric-learning models~\cite{cml1, cml2}.
Other studies proposed recommendation models that incorporate hyperbolic geometry into graph-based methods~\cite{Li2023_hyperbolic_rec,Sun2021_HGCF,hyper_rec_graph_1,hyper_rec_graph_2}; temporal point process~\cite{Zhou2023_hyperbolic_XRec}; or autoencoders~\cite{HAE2020}.
However, to the best of our knowledge, hyperbolic space has not been used on recommendation tasks where the visual data of items provides critical information, such as fashion item recommendation.

\vspace{-0mm}
\section{Methodology}
\vspace{-0mm}
The overview of our proposed model is illustrated in Figure~\ref{fig_structure}. The training data of our model is constructed as a set of triplets of users and positive/negative items, and is defined as follows:
\vspace{-1mm}
\begin{align}
\label{eq:dataset_rec}
D_{S} = \{(i,j,k)|i \in \mathcal{U}, j \in \mathcal{I}_{i}^{+}, k \in \mathcal{I} \backslash \mathcal{I}_{i}^{+} \},
\end{align}
where $\mathcal{U}$, $\mathcal{I}$, and $\mathcal{I}_{i}^{+}$ are the sets of users, items, and items purchased by the $i$-th user, respectively.

\vspace{-0mm}
\subsection{Training a Model in Hyperbolic Space}
Among several isometric models for modeling hyperbolic space, this paper adopts the Poincar\'{e} ball model, one of the most common models in computer vision~\cite{hyperbolic_cv_survey,Ge_2023_CVPR,cv_poincare_1,cv_poincare_2}.
In this paper, the induced geodesic distance between $\bm{x}$ and $\bm{y}$ in hyperbolic space $\mathbb{D}^{n}_{c} \coloneqq \{\bm{x}, \bm{y} \in \mathbb{R}^{n} : c \|\bm{x}\|^{2}, c \|\bm{y}\|^{2} < 1 \}$ for $c \geq 0$ is defined as follows:
\vspace{-1mm}
\begin{align}
\label{eq:hyperbolic_distance_gyrovector}
d_{c}(\bm{x},\bm{y}) \coloneqq \frac{2}{\sqrt{c}} \tanh^{-1} \Big(\sqrt{c} \| (-\bm{x}) \oplus_{c} \bm{y} \| \Big),
\end{align}
\vspace{-1mm}
\begin{align}
\label{eq:mobius_addition}
\bm{x} \oplus_{c} \bm{y} \coloneqq \frac{(1+2c \langle \bm{x},\bm{y} \rangle +c \| \bm{y} \|^{2})\bm{x} + (1-c \| \bm{x} \|^{2})\bm{y}}{1+2c \langle \bm{x},\bm{y} \rangle +c^{2} \| \bm{x} \|^{2} \| \bm{y} \|^{2}}.
\end{align}

\vspace{-1.5mm}
In addition, the following function maps a point $\bm{z} \in T_{\bm{q}}\mathbb{D}^{n}_{c}$ from a $n$-dimensional vector space $T_{\bm{q}}\mathbb{D}^{n}_{c}$, which is a tangent space of $\mathbb{D}^{n}_{c}$ at a point $\bm{q} \in \mathbb{D}^{n}_{c}$, to the hyperbolic space $\mathbb{D}^{n}_{c}$.
\vspace{-1.5mm}
\begin{align}
\label{eq:projection_euclidean2hyperbolic_space}
h(\bm{z}) = \exp^{c}_{\bm{q}}(\bm{z}) \coloneqq \bm{q} \oplus_{c} \Big( \tanh \Big( \sqrt{c} \frac{\lambda^{c}_{\bm{q}} \|\bm{z}\|^{2}}{2} \Big) \frac{\bm{z}}{\sqrt{c} \|\bm{z}\|} \Big),
\end{align}
\vspace{-0mm}
where $\lambda^{c}_{\bm{q}} = \frac{2}{1-c\|\bm{q}^2\|}$ is the conformal factor~\cite{NEURIPS2018_HperbolicNeuralNetwork}.
Detailed definitions can be found in the appendix or previous studies such as~\cite{Hyperbolic_Survey,hyperbolic_cv_survey}.

\vspace{-0mm}
\subsection{Neighbor-Attentive Aggregation}
Our proposed model extracts user preferences based on their purchase history and image data of each item.
Here, $\mathbf{U} = \{[\bm{u}_1; ...; \bm{u}_n; ...; \bm{u}_{N_u}] | \bm{u}_n \in \mathbb{R}^{D} \} \in \mathbb{R}^{N_u \times D}$ denotes the user embeddings, and $\mathbf{V} = \{[\bm{v}_1; ...; \bm{v}_n; ...; \bm{v}_{N_v}] | \bm{v}_n \in \mathbb{R}^{D} \} \in \mathbb{R}^{N_v \times D}$ and $\mathbf{P} = \{[\bm{p}_1; ...; \bm{p}_n; ...; \bm{p}_{N_v}] | \bm{p}_n \in \mathbb{R}^{D} \} \in \mathbb{R}^{N_v \times D}$ denote separate embeddings associated with the items. $N_u$ and $N_v$ are the numbers of users and items, respectively; $D$ is the embedding dimension; and $[\bm{a}; \bm{b}]$ denotes the concatenation operation between the vectors $\bm{a}$ and $\bm{b}$.

First, we calculate the embedding $\tilde{\bm{u}}_i$ that combines the $i$-th user embedding and item embeddings purchased by the user, inspired by~\cite{ACF}, as follows:
\vspace{-1mm}
\begin{align}
\label{eq:user_model}
\tilde{\bm{u}}_i = \frac{\bm{u}_i + \sum_{l \in A(\mathcal{I}_{i}^{+} \backslash j, L)} \alpha_{il} \bm{p}_{l}}{2},
\end{align}
where
$A(\mathcal{I}_{i}^{+} \backslash j, L)$ is an item set obtained by randomly sampling $L>0$ items from $\mathcal{I}_{i}^{+} \backslash j$.\footnote{If $L>|\mathcal{I}_{i}^{+} \backslash j|$, we retrieval all the items from $\mathcal{I}_{i}^{+} \backslash j$.}
Here, $\alpha_{il}$ represents the contribution (attention) score of the $l$-th item to the preference profile of the $i$-th user, and is defined as follows:
\vspace{-1mm}
\begin{align}
\label{eq:attention}
\hspace{-5mm} \alpha_{il} = \frac{\mathrm{exp}(\alpha'_{il} / \tau)}{\sum_{n \in A(\mathcal{I}_{i}^{+} \backslash j, L)} \mathrm{exp}(\alpha'_{in} / \tau)},
\end{align}
\vspace{-3mm}
\begin{align}
\label{eq:attention_base}
\alpha'_{il} =& \hspace{1.5mm} \bm{w}^{\top}_{2} \phi (\mathbf{W}_{\rm{u}} \bm{u}_i + 
\mathbf{W}_{\rm{v}} \bm{v}_l + \mathbf{W}_{\rm{p}} \bm{p}_l 
\notag \\&
+ \mathbf{W}_{\rm{f}} E(\bm{I}_l) + \bm{b}_1) + b_{2},
\end{align}
where $\tau$ is a scaling factor (the larger the value is, the sharper the attention score becomes), which we set $\tau=\sqrt{D}$, and $\phi(x) = \mathrm{max}(0, x)$ is the ReLU function. $\mathrm{E}(\cdot)$ is a pre-trained backbone encoder model and $\bm{I}_l$ is the $l$-th item's image data. $\mathbf{W}_{\rm{u}}, \mathbf{W}_{\rm{v}}, \mathbf{W}_{\rm{p}} \in \mathbb{R}^{D \times D}$, and $\mathbf{W}_{\rm{f}} \in \mathbb{R}^{D \times D_{\rm{pool}}}$ denote transformation matrices for the users, items, and item visual features extracted by the image encoder, respectively ($D_{\rm{pool}}$ is dimension size of the final pooling layer in the backbone model); $\bm{b}_{\rm{1}} \in \mathbb{R}^{D}$ is the bias vector of the first layer; and $\bm{w}_{\rm{2}} \in \mathbb{R}^{D}$ and $b_2 \in \mathbb{R}$ are the weighting vector and bias of the second layer.

\vspace{-0mm}
\subsection{Objective Function}
Following~\cite{Tran_HyperML}, the target parameters $\mathbf{\Theta}=\{\mathbf{U}, \mathbf{V}, \mathbf{P}, \\ \mathbf{W}_u, \mathbf{W}_v, \mathbf{W}_p, \mathbf{W}_f, \bm{b}_1, \bm{w}_{2}, b_{2}, \bm{q}\}$ are optimized as follows:
\vspace{-0mm}
\begin{align}
\label{eq:loss_hvabpr}
\argmin_{\mathbf{\Theta}} \mathcal{L} = \mathcal{L}_{\mathrm{hyp}} + \gamma \mathcal{L}_{\mathrm{adj}} + \lambda \| \mathbf{\Theta} \|^{2},
\end{align}
\vspace{-0mm}
\hspace{-1.2mm}where $\gamma>0$ is a scalar hyperparameter that balances the loss based on the hyperbolic distance ($\mathcal{L}_{\mathrm{hyp}}$) and the adjustment loss based on the Euclidean distance ($\mathcal{L}_{\mathrm{adj}}$), and $\lambda>0$ determines the degree of regularizing the model parameters ($\mathbf{\Theta}$).
Each loss function is defined as follows:
\vspace{-1mm}
\begin{align}
\label{eq:loss_hyp}
\hspace{-1mm} \mathcal{L}_{\mathrm{hyp}} \coloneqq \hspace{-3mm} \sum_{(i,j,k) \in D_{S}} \hspace{-2mm} \bigl[m + d_{c}^{2}(h(\tilde{\bm{u}}_{i}), h(\bm{v}_{j})) - d_{c}^{2}(h(\tilde{\bm{u}}_{i}), h(\bm{v}_{k})) \bigr]_{+},
\end{align}
\vspace{-2mm}
\begin{align}
\label{eq:loss_adj}
\hspace{-4mm} \mathcal{L}_{\mathrm{adj}} \coloneqq \sum_{(i,j) \in D_{S}} f_{\mathrm{adj}}(\tilde{\bm{u}}_{i}, \bm{v}_{j}) + \sum_{(i,k) \in D_{S}} f_{\mathrm{adj}}(\tilde{\bm{u}}_{i}, \bm{v}_{k}),
\end{align}
\vspace{-2mm}
\begin{align}
\label{eq:f_adj}
\hspace{-4mm} f_{\mathrm{adj}}(\bm{x}, \bm{y}) \coloneqq \Bigl[\frac{|d_{c}(h(\bm{x}), h(\bm{y})) - d_{\mathrm{euc}}(\bm{x},\bm{y})|}{d_{\mathrm{euc}}(\bm{x}, \bm{y})} \Bigr]_{+},
\end{align}
\vspace{-0mm}
\hspace{-1.3mm}where $[a]_{+}$ denotes $\max(0, a)$; $m>0$ is the margin in the hinge function; and $d_{\mathrm{euc}}(a, b)$ is the Euclidean distance.
During inference, the model recommends the items with the smallest hyperbolic distances to the user vector $\tilde{\bm{u}}_i$.

\begin{table}[t]
\centering
\caption{Statistics of the data sets used in our experiments on recommendation tasks.\label{tab:dataset_recommendation}}
\vspace{-3mm}
\scalebox{0.98}{
\begin{tabular}{lrrr}
\toprule
\textbf{Dataset} & \textbf{\#users} & \textbf{\#items} & \textbf{\#feedbacks} \\
\midrule
Amazon Women~\cite{amazonwomen_dataset}  & 30,631 & 285,269 & 849,801 \\
Amazon Men~\cite{amazonwomen_dataset}  & 27,892 & 120,758 & 451,932 \\
TOWN Women~\cite{app_zozotown}\footnotemark & 37,357 & 187,866 & 800,000 \\
\bottomrule
\end{tabular}
}
\end{table}
\footnotetext{ZOZOTOWN is the largest fashion e-commerce application in Japan.}

\begin{table}[t]
\centering
\caption{Comparison of our baselines; ``Vis'' and ``Hyp'' denote models with visual features and models in hyperbolic space, respectively. \label{tab:comparison_methods}}
\vspace{-3mm}
\scalebox{0.95}{
\begin{tabular}{lcc|lcc}
\toprule
\textbf{Method} & \textbf{Vis} & \textbf{Hyp} & \textbf{Method} & \textbf{Vis} & \textbf{Hyp} \\
\midrule
BPRMF~\cite{BPRMF} & \xmark & \xmark & VBPR~\cite{McAuley_VBPR} & \cmark & \xmark \\
HBPR~\cite{Tran_HyperBPR} & \xmark & \cmark & DVBPR~\cite{Kang2017_DVBPR} & \cmark & \xmark \\
HRec~\cite{hyper_rec_others_1} & \xmark & \cmark & DeepStyle~\cite{DeepStyle} & \cmark & \xmark \\
LFM~\cite{hyper_rec_others_2} & \xmark & \cmark & ACF~\cite{ACF} & \cmark & \xmark \\
HML~\cite{Tran_HyperML} & \xmark & \cmark & HVACF & \cmark & \cmark \\ \bottomrule
\end{tabular}
}
\end{table}

\vspace{-0mm}
\section{Experiments}
\vspace{-0mm}
\subsection{Experimental Settings}
\label{sec_experiment_setting}
We perform experiments using three data sets for fashion item recommendation tasks (shown in Table~\ref{tab:dataset_recommendation}).
Each data set is chronologically sorted and then divided into the train, validation, and test sets with a ratio of 7:1:2, respectively.
Besides, the ten previous models listed in Table~\ref{tab:comparison_methods} plus a random recommendation model are used for our baselines.
These baselines are chosen from the ``visually-aware recommendation models'' listed in \cite{McAuley2023_fashionrec_review} and the ``hyperbolic recommender systems'' listed in \cite{Hyperbolic_Survey}.
The area under the ROC curve (AUC) is used for evaluation, following previous work~\cite{BPRMF,McAuley_VBPR}.

\vspace{-0mm}
\begin{table}[t]
\centering
\caption{Results of our experiments (*DStyle = DeepStyle). The numbers denote the AUC scores of each baseline and our model. \label{tab:result_auc_test}}
\vspace{-3mm}
\scalebox{0.95}{
\begin{tabular}{lrrr}
\toprule
 \textbf{Method} & \textbf{Amazom (W)} & \textbf{Amazon (M)} & \textbf{TOWN (W)} \\
 \midrule
Rand      & 0.499~$\pm$0.002 & 0.499~$\pm$0.001 & 0.499~$\pm$0.002 \\ \midrule
BPRMF     & 0.702~$\pm$0.002 & 0.717~$\pm$0.003 & 0.741~$\pm$0.001 \\ \midrule
HBPR      & 0.711~$\pm$0.001 & 0.743~$\pm$0.001 & 0.716~$\pm$0.001 \\
HRec       & 0.762~$\pm$0.000 & 0.808~$\pm$0.002 & 0.784~$\pm$0.002 \\
LFM       & 0.679~$\pm$0.001 & 0.703~$\pm$0.002 & 0.685~$\pm$0.002 \\
HML       & 0.777~$\pm$0.002 & 0.818~$\pm$0.000 & 0.795~$\pm$0.001 \\ \midrule
VBPR      & 0.730~$\pm$0.000 & 0.756~$\pm$0.002 & 0.728~$\pm$0.001 \\
DVBPR     & 0.776~$\pm$0.001 & 0.818~$\pm$0.001 & 0.790~$\pm$0.000 \\
DStyle    & 0.728~$\pm$0.001 & 0.730~$\pm$0.004 & 0.700~$\pm$0.004 \\
ACF       & 0.767~$\pm$0.001 & 0.797~$\pm$0.002 & 0.772~$\pm$0.000 \\ \midrule
HVACF      & \textbf{0.804}~$\pm$0.001 & \textbf{0.830}~$\pm$0.002 & \textbf{0.810}~$\pm$0.001 \\
\bottomrule
\end{tabular}
}
\end{table}

\vspace{-0mm}
\begin{table}[t]
\centering
\caption{Results of our ablation studies. \label{tab:result_ablation}}
\vspace{-3mm}
\scalebox{0.8}{
\begin{tabular}{lrrr}
\toprule
\textbf{Variants} & \textbf{Amazon (W)} & \textbf{Amazon (M)} & \textbf{TOWN (W)} \\ \midrule
(complete) & \textbf{0.804}~$\pm$0.001 & \textbf{0.830}~$\pm$0.002 & \textbf{0.810}~$\pm$0.001 \\
\midrule
$d_c(\cdot) \Rightarrow d_{\mathrm{euc}}(\cdot)$ & 0.755~$\pm$0.001 & 0.790~$\pm$0.001 & 0.775~$\pm$0.002 \\
w/o $\mathcal{L}_{\mathrm{adj}}$ & 0.701~$\pm$0.001 & 0.749~$\pm$0.001 & 0.722~$\pm$0.001 \\
w/o aggregation & 0.745~$\pm$0.004 & 0.809~$\pm$0.002 & 0.779~$\pm$0.003 \\
w/o attention   & 0.798~$\pm$0.000 & 0.827~$\pm$0.001 & 0.807~$\pm$0.001 \\
attention w/o $E(\bm{I}_l)$ & 0.798~$\pm$0.001 & 0.826~$\pm$0.001 & 0.808~$\pm$0.001 \\
attention w/o $\bm{v}_l$ & 0.799~$\pm$0.001 & 0.828~$\pm$0.001 & 0.808~$\pm$0.001 \\
attention w/o $\bm{p}_l$ & 0.798~$\pm$0.000 & 0.827~$\pm$0.002 & 0.807~$\pm$0.001 \\
\bottomrule
\end{tabular}
}
\end{table}

\begin{figure}[t]
\begin{minipage}[b]{0.48\hsize}
\centering
\includegraphics[width=\textwidth]{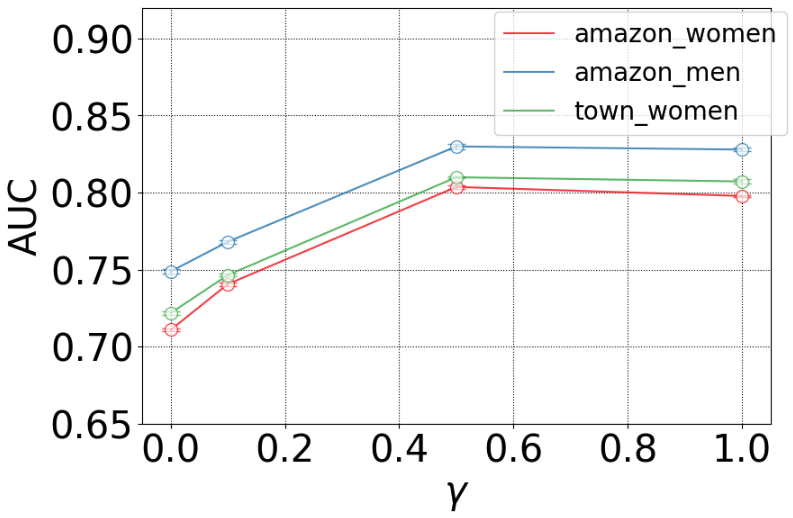}
\vspace{-5mm}
\subcaption{AUC w.r.t. balancing value $\gamma$ \label{fig_result_gamma}}
\end{minipage}
\begin{minipage}[b]{0.48\hsize}
\centering
\includegraphics[width=\textwidth]{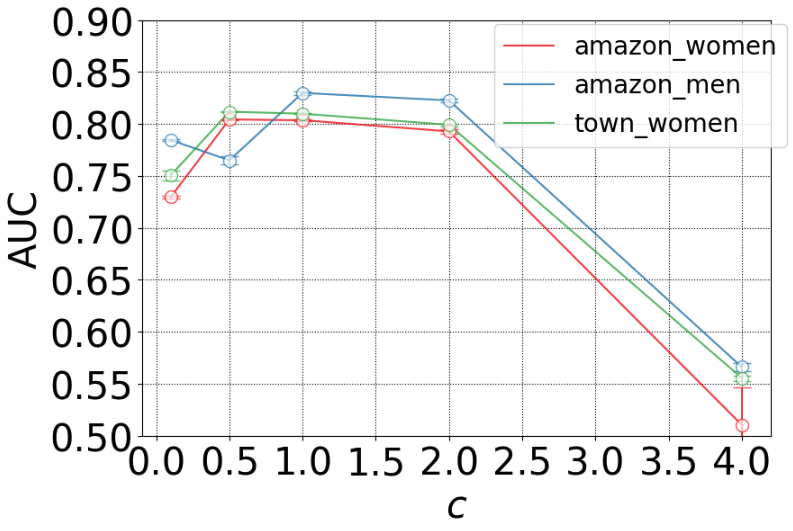}
\vspace{-5mm}
\subcaption{AUC w.r.t. scaling factor $c$ \label{fig_result_c}}
\end{minipage}
\vspace{-2mm}
\caption{Performance with different hyperparameters.}
\end{figure}

\begin{figure*}[t]
\begin{minipage}[b]{0.25\hsize}
    \centering
    \vspace{-0mm}
    \includegraphics[width=\textwidth]{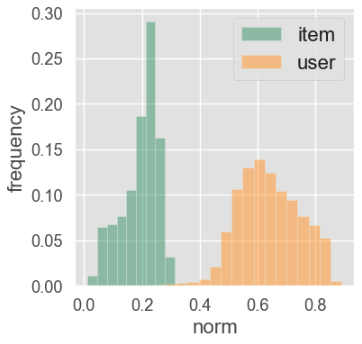}
    \vspace{-2.5mm}
    \caption{The embedding-norm distribution on Amazon Women. \label{fig_user_item_norm_hist_amazon_women}}
\end{minipage}
\begin{minipage}[b]{0.748\linewidth}
    \begin{minipage}[b]{0.328\linewidth}
    \centering
    \includegraphics[width=\textwidth]{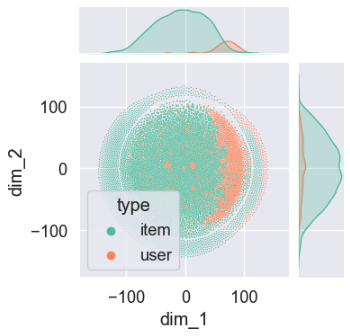}
    \vspace{-2mm}
    \subcaption{User and item representations \label{fig_user_item_mappping_amazon_women_epoch50}}
    \end{minipage}
    \begin{minipage}[b]{0.328\hsize}
    \centering
    \includegraphics[width=\textwidth]{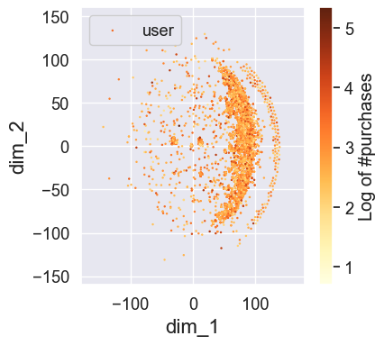}
    \vspace{-2mm}
    \subcaption{User representations \label{fig_user_mappping_amazon_women_epoch50}}
    \end{minipage}
    \begin{minipage}[b]{0.328\linewidth}
    \centering
    \includegraphics[width=\hsize]{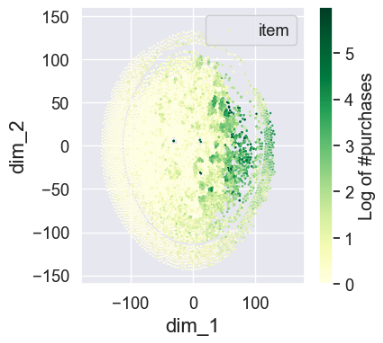}
    \vspace{-2mm}
    \subcaption{Item representations \label{fig_item_mappping_amazon_women_epoch50}}
    \end{minipage}
    \vspace{-2.5mm}
    \caption{User and item 2D-representations compressed by tSNE~\cite{Maaten2008_tSNE} on Amazon Women. \\ \,}
\end{minipage}
\end{figure*}

\vspace{-0.0mm}
\subsection{Implementation Details}
We use Riemannian Adam~\cite{radam} as the optimizer for HBPR, HRec, and LFM models based on the performance on the validation set, and use Adam~\cite{Adam} with decoupled weight decay~\cite{AdamW} for the other models.
We set the embedding dimension $D$ to $50$ and the batch size to $512$ for all models. We tune the learning rate $r \in \{0.01, 0.001, 0.0001\}$ and the hyperparameter for regularization $\lambda \in \{0.1, 0.01, 0.001\}$ for each model based on the performance on the validation set.
Inception V3~\cite{inception_v3} is used as the backbone image encoder model for all models with visual features, and its parameters are freezed during training.
For HVACF, we set $\gamma=0.5$, $c=1.0$, $L=32$, and $m=0.5$ based on the performance on the validation set.
We repeat each experiment three times with different random seeds, and report the mean and standard deviation values.

\vspace{-0mm}
\subsection{Results}
Table~\ref{tab:result_auc_test} shows the results. It demonstrates that our proposed model outperforms all baselines on three data sets. These results highlight the effectiveness of using hyperbolic space for fashion item recommender systems.

\vspace{-0mm}
\subsection{Ablation Studies}
We perform ablation studies and Table~\ref{tab:result_ablation} summarizes the results.
It shows that each component of our proposed model is indispensable to achieve high accuracy.
Notably, the performance of ``w/o $L_{\mathrm{adj}}$'' is lower than that of ``$d_c(\cdot) \Rightarrow d_{\mathrm{euc}}(\cdot)$'': a model trained in Euclidean space only.
The model ``w/o $\mathcal{L}_{\mathrm{adj}}$'' even underperforms baselines without visual features.
These results indicate the importance of considering both Euclidean and hyperbolic distances to train fashion item recommendation models.

\vspace{-0mm}
\subsection{Impacts of Hyperparameters}
Figure~\ref{fig_result_gamma} and \ref{fig_result_c} illustrate how the hyperparameters of our model affect its performance. Figure~\ref{fig_result_gamma} shows that setting $\gamma$ to a small value harms the performance, indicating that it is important to consider both the hyperbolic and Euclidean distances during training.
Figure~\ref{fig_result_c} shows that when $c$ is set too large, the performance drops sharply as a result of the radius of the embedding space (ball) $1/\sqrt{c}$ getting too small, which causes both users and items to be placed near the ball's surface during training, and results in inefficient training. 

\vspace{-0mm}
\subsection{Analysis of Embeddings}
\vspace{-0mm}
Figure~\ref{fig_user_item_norm_hist_amazon_women} shows the norms of the user and item embeddings trained on the Amazon Women data set in hyperbolic space (i.e., $h(\tilde{\bm{u}})$ and $h(\bm{v})$, respectively), which correspond to the distance distributions from the ball's origin of the space.
It clearly indicates that items and users are separated in the embedding space, with the items being placed near the origin and the users mapped near the ball's surface.

Next, we analyse the 2D-compressed representations of the users and items in
\textcolor{red}{4a}-\textcolor{red}{4c}, where each plot is colored based on the logarithmically transformed value of the number of the purchases or purchases made.
We can see that items with many purchases are mapped near the user embeddings.
We also calculate the correlation coefficient between the norms of item embeddings and the number of purchases made, and observe that it is $-0.360$.
This indicates that popular items tend to be mapped near the origin, and our proposed model separates items largely based on how popular they are among users.

\vspace{-0mm}
\section{Conclusion}
\vspace{-0mm}
In this paper, we proposed HVACF, a fashion item recommendation model trained in hyperbolic space.
Our experiments on three data sets showed that it performs better than previous models trained in Euclidean space, confirming the effectiveness of our model.
We also performed ablation studies and showed that it is crucial to consider both hyperbolic and Euclidean distances in the objective function.
Lastly, we analysed the trained embeddings of users and items and found that items and users are separated in hyperbolic space.
We also found that popular and unpopular items are also separated in the space, with the popular ones being centered near the origin.

{\small
}

\end{document}